\documentclass[11pt]{article}
\usepackage{url,epsfig}
\usepackage{geometry} 
\usepackage{sectsty} 
\usepackage[normalem]{ulem} 
\sectionfont{\sffamily\bfseries\upshape\large}
\subsectionfont{\sffamily\bfseries\upshape\normalsize} 
\subsubsectionfont{\sffamily\bfseries\upshape\normalsize} 
\makeatletter

\renewcommand\@seccntformat[1]{\csname the#1\endcsname.\quad}

\makeatother\renewcommand{\bibitem}{\vskip 2pt\par\hangindent\parindent\hskip-\parindent}

\makeatletter
\def\@maketitle{%
  \begin{center}%
  \let \footnote \thanks
    {\large \@title \par}%
    {\normalsize
      \begin{tabular}[t]{c}%
        \@author
      \end{tabular}\par}%
    {\small \@date}%
  \end{center}%
}
\makeatother

\title{\bf Convincing Evidence\footnote{For a volume on theoretical or methodological research on authorship, functional roles, reputation, and credibility on social media.  We thank Paul Rosenbaum for helpful comments, Sorin Matei and Elisa Bertino for inviting this article and the National Science Foundation for partial support of this work.}\vspace{.1in}}
\author{Andrew Gelman\footnote{Department of Statistics, Columbia University, New York, N.Y.} \and
Keith O'Rourke\footnote{O'Rourke Consulting, Ottawa, Ontario}}
\date{25 June 2013}
\begin{document}
\maketitle
\thispagestyle{empty}

\begin{abstract}
Textbooks on statistics emphasize care and precision, via concepts such as reliability and validity in measurement, random sampling and treatment assignment in data collection, and causal identification and bias in estimation.  But how do researchers decide what to believe and what to trust when choosing which statistical methods to use? How do they decide the credibility of methods? Statisticians and statistical practitioners seem to rely on a sense of anecdotal evidence based on personal experience and on the attitudes of trusted colleagues.  Authorship, reputation, and past experience are thus central to decisions about statistical procedures.
\end{abstract}

The rules of evidence as presented in statistics textbooks are not the same as the informal criteria that statisticians and practitioners use in deciding what methods to use.

According to the official rules, statistical decisions should be based on careful design of data collection, reliable and valid measurement, and something approximating unbiased or calibrated estimation.  The first allows both some choice of the assumptions and an opportunity to increase their credibility, the second tries to avoid avoidable noise and error and third tries to restrict to methods that are seemingly fair. This may be fine for evaluating psychological experiments, or medical treatments, or economic policies, but we as statisticians do not generally follow these rules when considering improvements in our teaching (Gelman and Loken, 2012) nor when deciding what statistical methods to use.

Did Fisher decide to use maximum likelihood because he evaluated its performance and the method had a high likelihood?  Did Neyman decide to accept a hypothesis testing framework for statistics because it was not rejected at a 5\% level?  Did Jeffreys use probability calculations to determine there were high posterior odds of Bayesian inference being correct?  Did Tukey perform a multiple comparisons analysis to evaluate the effectiveness of his multiple comparisons procedure?  Did Rubin use matching and regression to analyze the efficacy of the potential-outcome framework for causal inference?  Did Efron perform a bootstrap of existing statistical analyses to demonstrate the empirical effectiveness of resampling?  Do the authors of textbooks on experimental design use their principles to decide what to put in their books?  No, no, no, no, no, no, and no.  We do know some psychometricians who fit item response models to evaluate their exam questions, and this is one of the very few examples we can think of where statistics researchers are using statistical principles to make professional decisions. Gigerenzer and colleagues have done some for deciding to use percentages versus natural frequencies for better understanding of analysis when those doing the analysis are medical students or faculty.

How, then, do we gain our knowledge about how to analyze data?  This is a question that arises over and over as we encounter new sources of data that are larger and more structured than ever before.  How we decide to believe in the effectiveness of a statistical method? Following Gelman (2013), here are a few potential sources of evidence:
\begin{enumerate}
\item Mathematical theory (for example, coherence of inference or asymptotic convergence);
\item Computer simulations (for example, demonstrating approximate coverage of interval estimates under some range of deviations from an assumed model);
\item Solutions to toy problems (for example, the comparison of Rubin (1981) of a partial pooling estimate for a test-preparation program in eight schools to the no pooling or complete pooling estimates);
\item Improved performance on benchmark problems (for example, getting better predictions for the Boston Housing Data (Harrison and Rubinfeld, 1978), an example much beloved of textbook writers in statistics and computer science);
\item Cross-validation and external validation of predictions (see Vehtari and Ojanen, 2012), as can be done in various examples ranging from education to business to election forecasting;
\item Success as recognized in a field of application (for example, a statistical method that is used and respected by biologists, or economists, or political scientists);
\item Success in the marketplace of software or textbooks (under the theory that if people are willing to pay for something, it is likely to have something to offer);
\item Face validity:  whether the method seems reasonable.  This can be a minimum requirement for considering a new method.
\end{enumerate}

As noted by Gelman (2013), ``None of these is enough on its own. Theory and simulations are only as good as their assumptions; results from toy problems and benchmarks don't necessarily generalize to applications of interest; cross-validation and external validation can work for some sorts of predictions but not others; and subject-matter experts and paying customers can be fooled.
The very imperfections of each of these sorts of evidence and how they apply to different user populations and settings gives a clue as to why it makes sense to care about all of them. We can't know for sure so it makes sense to have many ways of knowing.''  Informal heuristic reasoning is important even in pure mathematics (Polya, 1941). 

There is also the concern that a statistical method will be used differently in the field than in the lab, so to speak---or, to give this problem a pharmaceutical spin, that a new method, approved for some particular class of problems, will be used ``off-label'' in some other setting. Rubin (1984) discusses concerns of recommending methods of analysis for repeated use by those (and often ourselves) with limited statistical expertise, limited resources, and limited time. To further complicate this, there has long been experimental evidence that optimal methods of information processing do not always not lead to optimal human performance, and this varies by level of skill, incentives and time pressure (Driver and Streufert, 1967).

We may also wish to consider how we should choose between methods in given applications for ourselves, to recommend to colleagues of similar or different levels of technical skill, and to communities of users who are not full-time statisticians or quantitative analysts. That is, how should we go about approving statistical methods for use in various applications by various users, making reasoned, critical choices. To do this we lean on background material, again following the model of the choice of medical treatments for use by various professionals or end users. Our primary objective is to maximize the rate of learning about the empirical application while minimizing the rate and magnitude of mistakes. These goals require a sort of meta-evidence that is not captured by any single sort of inquiry.  More generally, we have argued that stories, to the extent that they are anomalous and immutable, are central to building understanding in social science (Gelman and Basb$\o$ll, 2013).

How do we build trust in statistical methods and statistical results?  There are lots of examples but out of familiarity we will start with my (Gelman's) own career. My most cited publications are my books and my methods papers, but I think that much of my credibility as a statistical researcher comes from my applied work. It somehow matters, I think, when judging my statistical work, that I've done (and continue to do) real research in social and environmental science. 

Why is this? It's not just that my applied work gives me good examples for my textbooks. It's also that the applied work motivated the new methods. Most of the successful theory and methods that my collaborators and I have developed, we developed in the context of trying to solve active applied problems. The methods have faced real challenges and likely have been appropriately tempered in some generally relevant ways. At the very least, any of our new methods are computationally feasible, have face validity, and solve at least one applied problem.  In developing and applying these methods, we weren't trying to shave a half a point off the predictive error in the Boston housing data; rather, we were attacking new problems that we couldn't solve in any reasonable way using existing approaches.

That's fine, but in that case who cares if the applied work is any good? To put it another way, suppose my new and useful methods had been developed in the context of crappy research projects where nobody gave a damn about the results? The methods wouldn't be any worse, right?
The mathematics does not care whether the numbers are real or fake. I have an answer to this one: If nobody cared about our results we would have little motivation to improve. Here's an example. A few years ago I posted some maps based on multilevel regression and post stratification of pre-election polls to show how different groups of white people voted in 2008. The blogger and political activist Kos noticed that some of the numbers in my maps didn't make sense. Kos wasn't very polite in pointing out my mistakes, but he was right. So I want back and improved the model (with the collaboration of my student Yair Ghitza). It took a few months, but at the end we had better maps---and also a better method (which was later published in the {\em American Journal of Political Science}). This all only happened because I and others cared about the results. Kos and other outsiders performed severe testing of our conclusions, which ruined the face validity of our claims, and then we went through a process of trying getting the representation (model) less wrong for this particular empirical problem. If all we were doing was trying to minimize mean squared predictive error, I doubt the improvements would've done anything at all.  Indeed, it turns out that it is difficult to identify improvements in hierarchical models via cross-validated mean squared error, even with large sample sizes (see Wang and Gelman, 2013, a paper that was developed as our attempt to understand our challenges in comparing different models for this and similar small-area estimation problems).

This is not to say that excellent and innovative statistical theory can't be developed in the absence of applications or, for that matter, in the context of shallow applications. For example, my popular paper on prior distributions for group-level variance parameters (Gelman, 2006) came through my repeated study of the 8-schools problem of Rubin (1981), a dead horse if there ever was one.  Hierarchical modeling is still unsettled, and it was possible to make a contribution in this field using an example without any current applied interest. In many cases, though, seriousness of the application, the focus required to get details right (the representation less wrong), was what made it all work.

As suggested earlier, one possible way to better cover all the issues and challenges of deciding on which statistical method is most credible is to borrow from the regulatory review and approval perspective that is brought to bear on deciding which medical treatments should be approved, for what purpose, by whom, in which circumscribed situations. The actual reasons for statisticians and non-statisticians choosing statistical methods in practice may be less interesting, likely being largely more to do with perceived authority, sociology (peer group), psychology, and economics (skill level, access to software, budget limitations of client) than any putative evidence. A regulatory perspective would attempt to set this aside, at least until the benefits and harms of methods have been assessed along with their relative value/importance and the real uncertainties about these are clarified.  We are not literally suggesting that statistical methods be subject to regulatory approval but rather that this perspective can help us make sense of the mix of information available about the effectiveness of different research methods.

More recently, it has been argued that truth comes form ``big data'' (see Hardy, 2013, for a contrary view).  We agree with the saying ``data trumps analysis'' but in practice it can be easier to work with small datasets we understand than with large datasets with unknown selection biases. 
For example, in our analysis of home radon levels (Lin et al., 1999), we used national and state-level random sample surveys of about 80,000 homes (which sounds like a lot but is not that much when you consider that radon concentrations vary spatially, and there are over 3000 counties in the United States), ignoring millions of measurements that were collected by individual homeowners, buyers, and sellers, outside of any survey context.  We suspect that were we to have ready access to the large set of self-selected data, it would be possible to perform some analysis to calibrate with respect to the more carefully gathered measurements and get the best of both worlds.  In that particular example, however, we doubt this will ever happen because there is not such a sense of urgency about the problem; our impression is that just about everyone who is worried about radon has already had their house measured.  For other problems such as medical treatments (or, in the business world, social advertising), we suspect that much can be learned (indeed, is already being learned) by combining experimental measurements with larger available observational data (see, for example, Kaizar, 2011, and Chen, Owen, and Shi, 2013).

This article is appearing in a volume whose goal is to consider ``a future agenda for theoretical or methodological research on authorship, functional roles, reputation, and credibility on social media.''  We do not have a clear sense of what this agenda should be, but we think it important to recognize the disconnect between our official and unofficial modes of reasoning in statistics, and the many different sources of practical evidence we use to make professional decisions.  A fruitful direction of future research could be the formalization of some of our informal rules, much in the way that Rosenbaum (2010) formalized and critiqued the well-known rules of Hill (1965) in epidemiology.

\section*{References}

\noindent

\bibitem Chen, A., Owen, A. B., and Shi, M. (2013).  Data enriched linear regression.\\{\tt http://arxiv.org/abs/1304.1837}

\bibitem Driver, M. J., and Streufert, S. (1969).  Integrative complexity:  an approach to individuals and groups as information-processing systems.  {\em Administrative Science Quarterly} {\bf 14}, 272--285.

\bibitem Gelman, A. (2006). Prior distributions for variance parameters in hierarchical models. {\em Bayesian Analysis} {\bf 1}, 515--533.

\bibitem Gelman, A. (2013).    How do we choose our default methods?  For the Committee of Presidents of Statistical Societies 50th anniversary volume.

\bibitem Gelman, A., and Basb$\o$ll, T. (2013).  When do stories work? Evidence and illustration in the social sciences.  Technical report, Department of Statistics, Columbia University.
  
\bibitem Ghitza, Y., and Gelman, A. (2013).  Deep interactions with MRP: Election turnout and voting patterns among small electoral subgroups. {\em American Journal of Political Science}. 

\bibitem Gelman, A., and Loken, E. (2012).  Statisticians: When we teach, we don't practice what we preach. {\em Chance} {\bf 25} (1), 47--48.

\bibitem Hardy, Q. (2013).  Why big data is not truth.  {\em New York Times}.\\
{\tt http://bits.blogs.nytimes.com/2013/06/01/why-big-data-is-not-truth/}

\bibitem Harrison, D., and Rubinfeld, D. L. (1978).  Hedonic prices and the demand for clean air.  {\em Journal of Environmental Economics and Management} {\bf 5}, 81--102.

\bibitem Hill, A. B. (1965).  The environment and disease:  association or causation? {\em Proceedings of the Royal Society of Medicine} {\bf 58}. 295--300.

\bibitem Kaizar, E. E. (2011).  Estimating treatment effect via simple cross design synthesis.  {\em Statistics in Medicine},
{\bf 30}, 2986--3009.

\bibitem Lin, C. Y., Gelman, A., Price, P. N., and Krantz, D. H. (1999). Analysis of local decisions using hierarchical modeling, applied to home radon measurement and remediation (with discussion). {\em Statistical Science}, {\bf 14}, 305--337.

\bibitem Polya, G. (1941).  Heuristic reasoning and the theory of probability.  {\em American Mathematical Monthly} {\bf 48}, 450--465.

\bibitem Rosenbaum, P. R. (2010).  {\em Observational Studies}, second edition.  New York:  Springer.

\bibitem Rubin, D. B. (1981).  Estimation in parallel randomized experiments.
{\em Journal of Educational Statistics} {\bf 6}, 377--401.

\bibitem Rubin, D. B. (1984).  Bayesianly justifiable and relevant frequency
calculations for the applied statistician.  {\em Annals of Statistics}
{\bf 12}, 1151--1172.

\bibitem Vehtari, A., and Ojanen, J. (2012). A survey of Bayesian predictive methods for model assessment,
selection and comparison. {\em Statistics Surveys} {\bf 6}, 142--228.

\bibitem Wang, W., and Gelman, A. (2013).  Cross-validation for hierarchical models.   Technical report, Department of Statistics, Columbia University.

\end{document}